# Characterization of Anomalous Pair Currents in Josephson Junction Networks

**I. Ottaviani, M. Lucci, R. Menditto, V. Merlo, M. Salvato (\*), and M. Cirillo (\*)**

*MINAS-Dipartimento di Fisica*, Università di Roma *Tor Vergata*, I-00173 Roma, Italy

**F. Müller and T. Weimann**

*Physikalisch Technische Bundesanstalt*, Bundesallee 100, D-38116 Braunschweig, Germany

**M. G. Castellano, F. Chiarello, and G. Torrioli**

*IFN-CNR*, Via Cineto Romano, I-00156 Roma, Italy

**R. Russo**

*Istituto di Cibernetica del CNR "E. Caianiello"*
Via Campi Flegrei 34
I-80078, Pozzuoli, Italy

(\*) Also CNR-SPIN Institute, Italy

## Abstract

Measurements performed on superconductive networks shaped in the form of planar graphs display anomalously large currents when specific branches are biased. The temperature dependencies of these currents evidence that their origin is due to Cooper pair hopping through the Josephson junctions connecting the superconductive islands of the array. The experimental data are discussed in terms of a theoretical model which predicts, for the system under consideration, an inhomogeneous Cooper pair distribution on the superconductive islands of the network.

Patterns of superconductive islands connected by Josephson junctions (often referred to as Josephson junctions networks or arrays) have attracted noticeable interest in the past decades both from the fundamental [1] and applied physics [2] point of view. An inhomogeneous distribution of Cooper pairs on a Josephson network in a "classical" regime, in which the Josephson energy dominates over the charging energy of the junctions, was investigated by Burioni et al. [3,4]. Their theoretical analysis showed that in Josephson networks shaped in the form of a symmetric comb, or other graph-like structures, the hopping between neighbour superconducting islands generates a nonuniform distribution of bosons, physically represented by Cooper pairs, along the branches. A thorough mathematical formulation of this problem has been recently published [5] and measurements have pointed toward a remarkable qualitative agreement of this theoretical model with experimental reality [6,7]. We have undertaken a systematic analysis of inhomogeneous distribution of Cooper pairs on a comb graph by measuring the current-voltage characteristics (*I-V* curves) of specific arrays and determining temperature and magnetic field-induced dependencies of the curves. The collected data allow a quantitative analysis which can be compared with the theoretical expectations.

Our experiments are made on arrays of high quality Josephson tunnel junctions based on the niobium trilayers (Nb-AlOx-Nb) technology [8]; for these specific experiments, in order to minimize the scatter in the geometrical parameters of the junctions, a hybrid electron beam-optical technology was used and, in particular, the areas of the tunnel junctions were defined by electron beam lithography. In Fig. 1a we show an optical microscope image zooming a region of a sample and the inset shows schematically a symmetric comb graph where the dots represent the superconductive islands and the lines are the connections between these for which Josephson junctions are responsible. The central pattern of the structure on which every island is connected to four neighbours through Josephson junctions generates what we call backbone array (BA), while the lines departing perpendicularly from it give rise to what we call fingers arrays (FA). The particular of Fig. 1a shows four cross-shaped superconducting islands of the BA and eight FA islands departing from these; some islands are generated by the base electrode (those like A and D) and others (like B and C) by the wiring electrode [8]; the dashed rectangles indicate where the Josephson junctions are located. The samples that we fabricated contain 400 backbone superconducting islands and 400 finger arrays; the high number of fingers was designed in the attempt to approach the thermodynamic limit of the number of bosons considered by the theory [3,4]. It is worth noting that the present design contains a relevant correction with respect to previously fabricated samples [6,7]: the BA was now designed in a way to meet more closely the requirements of a symmetric comb topology. In fact we see in Fig. 1a that each superconducting island of the backbone has four first nearest neighbour islands, as it should be (see inset), while in the previous experiments half of the islands of the BA had only two first nearest neighbour islands. Each island of the fingers, instead, is connected only to two first nearest neighbour islands. This difference in the number of neighbors revealed to be one of the main causes of the topology-induced condensation phenomenon on graphs [3,4].

A main concern when dealing with the theory reported in refs. 3 and 4 is the fact that the calculated effects (nonuniform distribution of Cooper pairs on the islands) should be observable when the Josephson coupling energy $E_j$ [9] between the islands is of the order of the thermal energy, namely when $E_j = \phi_0 I_c/2\pi \cong k_B T$ (see equation 18.3 of ref. 6), where $I_c$ is the maximum Josephson pair current, $\phi_0 = 2.07 \times 10^{-15}$ Wb is the flux-quantum and $k_B = 1.38 \times 10^{-23}$ J/K is the Boltzmann constant. We have to bear in mind that our bosons exist only below *9.2 K*, the superconducting transition temperature of niobium, and therefore in order to preserve the physical sense of the theory, the Josephson energy must be adequately tuned. For a maximum pair current of *1 μA*, the Josephson energy would be *3x10$^{-22}$ J* and comparable with the thermal energy which, in the *(1-9)K* temperature interval ranges in the *(10$^{-22}$-10$^{-23}$)J* interval. We conclude that a critical current of the order of few microamperes could be appropriate for the experimental observations [10]; for this reason indeed in previous investigations [6,7] the interval *(1-10)μA* for the maximum critical currents of the junctions was targeted.

We must point out, however, that when measurements are performed biasing Josephson junctions with an external current $I$ (as in our case) a relevant energy to be compared with thermal excitations is not the bare $E_j$, but rather the height of the Josephson potential

$$\Delta U = 2E_j \left[ \sqrt{1-(I/I_c)} - \frac{I}{I_c} \cos^{-1}(\frac{I}{I_c}) \right]$$

[9]. In Fig. 1b we plot $\Delta U$ as a function of $I$ for different values of the maximum Josephson current $I_c$; in the figure the horizontal straight line represents $k_B T$ for $T=9K$. We see that for high values of the bias currents, for all the curves, $\Delta U$ decreases below the thermal threshold; since when measuring current-voltage characteristics of specific branches-arrays we trace switches from the highest values of the Josephson, this experimental technique is such that along current-biased branches a lowering of the Josephson potential is generated. We do not exclude that this technique eases the observation of the theoretically predicted nonuniform distribution of bosons along the biased branches.

In Fig. 2a we show the current-voltage characteristics of the arrays of the sample CB428. The characteristics are traced biasing the arrays through four contacts pads placed at their ends, as shown in the inset of Fig. 1a. The *I-V* curves refer to three arrays and we notice that switching current distributions and subgap resistance indicate very good quality and uniformity of the samples. We have expanded the regions of the characteristics indicated by the zoom-squares 1 and 2 in Fig. 2a respectively in Fig. 2b and Fig. 2c. In the zoom of Fig. 2b we see the Josephson current distributions of BA and FA and we observe that the currents of BA are roughly *0.5 μA* above those of the FA. In this figure we also show the currents of two "reference" arrays which are two arrays having the same geometry of the backbone and finger arrays but are not embedded in the graph structure and are isolated from it. We call these reference backbone array (RBA) and reference finger array (RFA), respectively: we note that the two reference arrays have current amplitudes equal within the experimental error, while BA and FA have currents higher respectively of *1μA* and *0.5 μA* than the reference arrays. The differences between "comb-embedded" and "reference" arrays were pointed out for the first time in ref. 6 and were well characterized in ref. 7, however now, due to fact that in our new design the BA is "topologically" consistent with the theoretical model we record very uniform currents for all the backbone junctions at all temperatures, which was not the case in ref. 7 (see Fig. 2b of the paper). We remark that in the present case the areas of the junctions have a very limited spread (about *1%*), therefore geometrical factors can be safely ruled out as cause of the observed differences in critical currents of the arrays: indeed we see that the reference arrays have currents identical within experimental uncertainty. Differences exist between comb-branches arrays (BA and FA) and their counterpart reference arrays and these cannot be ascribed to experimental and design uncertainties.

As illustrated in Fig. 2b, the currents of the FA are slightly lower than the currents of the BA, a condition which is observed for most of the junctions of the arrays; in Fig. 2c, however, we see that close to the gap sum voltage the currents of the finger array become substantially higher than the currents of the backbone. We also show in Fig. 2d that this anomalous current increases substantially lowering the temperature; we remark that for sample CB428 apparently only two junctions of the FA exhibit an anomalous large current which, at *500 mK* is of the order of *4μA* corresponding to *25%* of the Josephson critical current. We have observed the anomalous increase of finger currents close to the gap sum voltage on several samples and we notice that this effect was also visible on the experiments reported in ref. 7 (see Fig. 3 of the paper), but in that work the phenomenon was not mentioned; a similar effect has also been measured on star-shaped graph arrays [11]. In Fig. 3a we show a striking example of anomalous currents in the finger array for the sample CB422 having a current density of *480 A/cm²* and a maximum Josephson current $I_c$=*43μA* at *4.2K*: the specific record of Fig. 3a was taken at a temperature of *1K*, and we see in this case that the anomalous extra current almost reaches the level where the normal state resistance takes over in the conduction process above the gap. At this temperature the span of the anomalous current above the average Josephson critical currents is *27 μA* and this corresponds to being *56%* of the critical currents of the arrays, more than twice the percentage observed for the sample CB428.

We have systematically investigated the response of the anomalous excess currents to temperature variations. The amplitude of these currents is obtained as the difference between a maximum and a minimum defined as follows: the maximal current is determined by the value where the last *2.5mV* switch before the gap-sum occurs, and the minimum ($I_{cmin}$) is the extrapolated value (from the main array switching line) of the Josephson current at the corresponding voltage. The results that we obtain for the specific case of the sample CB422 are shown in Fig. 3b; in this plot we also show the dependence upon temperature of $I_{cmin}$ (which is essentially the Josephson current). Both the measured temperature dependencies have been fitted, see the lines through the data in the figure, with the Ambegaokar-Baratoff functional relationship [9] in which we approximated the gap by the equation $\Delta(T) = \Delta(0)\tanh(A\sqrt{\frac{T_c - T}{T}})$ [12] where *A* is a constant. For the fits of Fig. 3b we have *Tc=7.28K* and *A=1.45* for the anomalous current (squares) and *Tc=8.47K* and *A=1.3* for the Josephson current (triangles). Since the only significant difference in the fitting of Fig. 3b is the value of the condensation temperature $T_c$ we speculate that the two phenomena are relative to different condensation processes in which Cooper pairs are involved. It is worth noting that the "anomalous" currents appear at a temperature of *7.28K* which is *1.2K* below the superconducting condensation temperature of the samples. In other words, below the BCS condensation temperature an additional condensation occurs at a lower characteristic temperatures but this phenomenon has nothing to do with the birth of the Cooper pairs: according to the theoretical model it is generated by a nonuniform distribution of pairs on the islands.

  The data of Fig. 3b show that the extra tunnelling currents observed on the finger arrays have a pair-current nature whose temperature dependence is regulated by a BCS functional relation. Moreover, since the Ambegaokar-Baratoff relationship gives, close to the critical temperature, a dependence of the anomalous current upon temperature like (*1- T/T$_c$*) we recover, in this limit, a prediction of Burioni et al [3]. In Fig.4a we plotted for different temperatures, the amplitude of the anomalous current *vs* the maximum Josephson current for CB422 and another sample, CB427 having a current density of *200 A/cm²*. In the figure we report, for different temperatures, the maximum Josephson current on the horizontal axis and the amplitude of the anomalous currents on the vertical axis : the two straight line dependencies indicate that the ratio of the physical quantities is a constant not depending upon temperature. If we assume that the observed anomalous currents are generated by boson hopping between the islands and associate it with the "filling factor" of the theory [3,4] we recover here a theoretical prediction, namely that the ratio between filling factor and Josephson energy is a constant not dependent upon temperature [3,4].

The theory [3,4] also indicates that the population of bosons on the backbone islands is maximal and therefore one would expect that the currents of the finger array could reach the value of the backbone currents: the FA shares one superconductive island (see Fig. 1a) with the backbone and two junctions of the fingers could have value of the current close to that of the backbone. Instead, in Fig. 2 and Fig. 3 we see FA Josephson currents having currents substantially higher than all the backbone currents. This phenomenon can be interpreted as follows: the total current flowing between islands of the arrays along one direction can be written as $\frac{dq_i}{dt} = \frac{\partial q_i}{\partial t} + \frac{\partial q_i}{\partial x_i}\frac{dx_i}{dt}$, where $q_i$ represents the total charge on each superconductive island, $\frac{\partial q_i}{\partial x_i}$ the charge gradient between neighbouring islands and $\frac{dx_i}{dt}$ the speed of the charge flow. As far as the BA is concerned, according to the theoretical model, the second term on the right hand side of this equation is zero since the charge distribution on all the BA islands is uniform [3,4]. On the FA instead, the islands close to the backbone experience a variation of charge carriers density and therefore the spatial derivative of the charge $\frac{\partial q_i}{\partial x_i}$, performed where the FA "crosses" the backbone, can give a relevant contribution to the total current. Since ref. 3 and 4 report the specific spatial - dependency of the density of bosons on the islands of the FA, the spatial derivative of the current(s) due to Cooper pairs hopping between the islands can be calculated, resulting in a exponential decay. It is worth noting that on reference arrays there are no effects that one could ascribe, as above, to charge non uniformities: reference arrays present uniform Josephson currents and related standard properties.

We remark now that the decrease of current in Fig. 3 takes place in correspondence of voltage steps which are, on the average, of the order of *5 mV* meaning that two junctions of the FA share roughly the same current (as it was the case in Fig. 2 for sample CB428): this result is consistent with the fact that the FA that we are measuring is symmetric with respect to the backbone line, therefore each current step corresponds to a jump generated by two junctions situated at the same distance from the BA. The dependence that we extracted from the samples CB422 and CB427 is shown in the semi-log plot of Fig. 4b where we see that the functional form is a decreasing exponential. Since we do not have access to all the junctions of the arrays, in principle, we cannot distinguish whether the currents of the pairs of junctions originating the results of Fig. 3 correspond to pairs located at the two sides of the BA and at the same distance from it. However, as "singular" (and centre of symmetry) points of the FA we can identify only the one crossing the BA : the ends in fact generate no effects on the Josephson current distribution in other arrays (BA and reference arrays) and there is no reason why the FA should behave differently. We believe at this point that the exponential decay shown in Fig. 4b is generated by the predicted decrease of population of bosons [3.4] when moving from backbone islands toward the end of the fingers.

In conclusion, we have reported evidence that topology-induced effects in arrays of Josephson junctions can generate relevant gradients in the distribution of Cooper pairs on the superconductive islands of the arrays. Our results do not contradict previous experimental observations [6,7] and provide further quantitative input toward a full characterization of this intriguing phenomenon.

We acknowledge fruitful discussions with Raffaella Burioni, Davide Cassi, and Francesco Fidaleo. Our interest in the topic of the Josephson graph-arrays was stimulated by Mario Rasetti and Pasquale Sodano and we wish to mention their role in originating the work herein presented.

**Figure Captions**

Fig. 1. a) Optical microscope image of a portion of our comb-shaped array : in the inset the dots are the superconductive islands. In the main panel (A and B)-like cross-shaped features are backbone islands; (C and D)-like islands belong to fingers. The dashed areas indicate junctions location; b) the height of the washboard potential of the Josephson junctions plotted as a function of the external bias current for different values of the maximum critical current $I_c$. When the potential curves fall below the straight horizontal like, corresponding to a temperature of *9K* , thermal excitations compete with Josephson potential energy.

Fig. 2. a) Current-voltage characteristics of three arrays of the sample CB428 at *4.2 K*. The expansion of the areas indicated by the zooming squares 1 and 2 are reported respectively in b) and c). In b) we have added the currents of RFA array to compare it with the RBA. In d) we show the feature observed in c) for different values of the temperature.

Fig. 3. a) The anomalous excess current at the gap-sum voltage of the sample CB422. We see that the FA (finger array) currents become much greater than the BA (backbone array currents). b) dependence of the excess current upon temperature (squares) and of $I_{cmin}$ (the maximum Josephson current). The curves fitting the data are obtained from the Ambegaokar-Baratoff (AB) functional relationship.

Fig. 4. a) The anomalous currents plotted vs. Josephson currents: each point corresponds to a different temperature; b) decay of the anomalous excess current as a function of the voltage distance from the gap-sum voltage for two samples (CB427 and CB422). The straight line behavior in this semi-log plot indicates and exponential functional dependence.

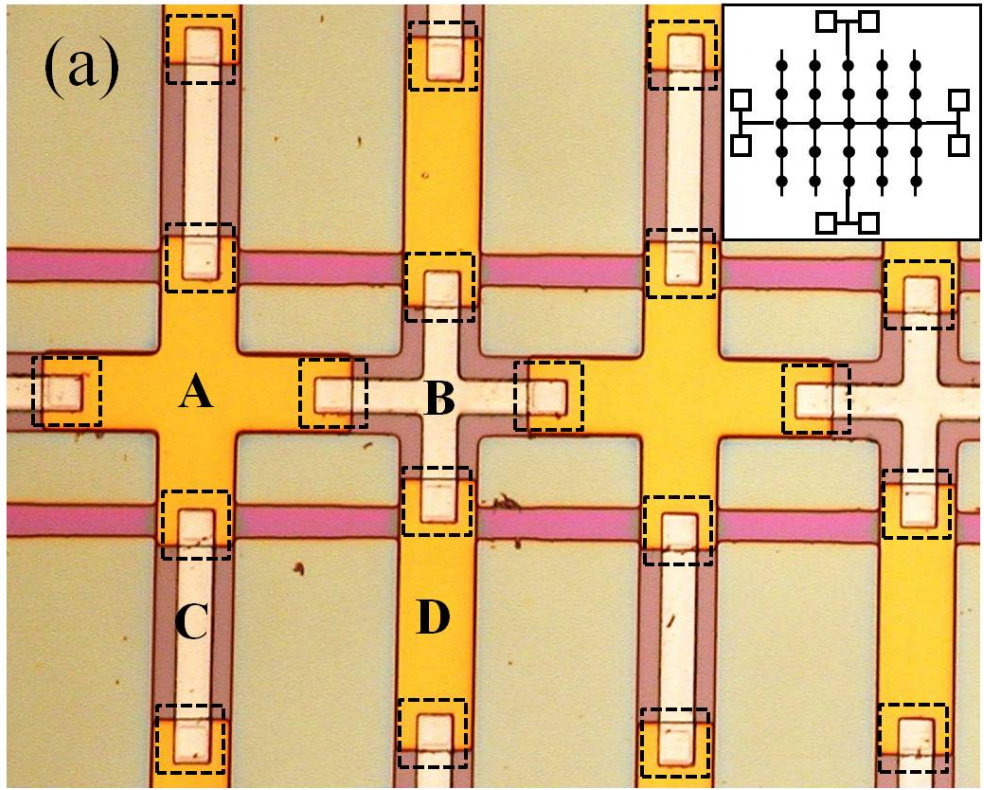

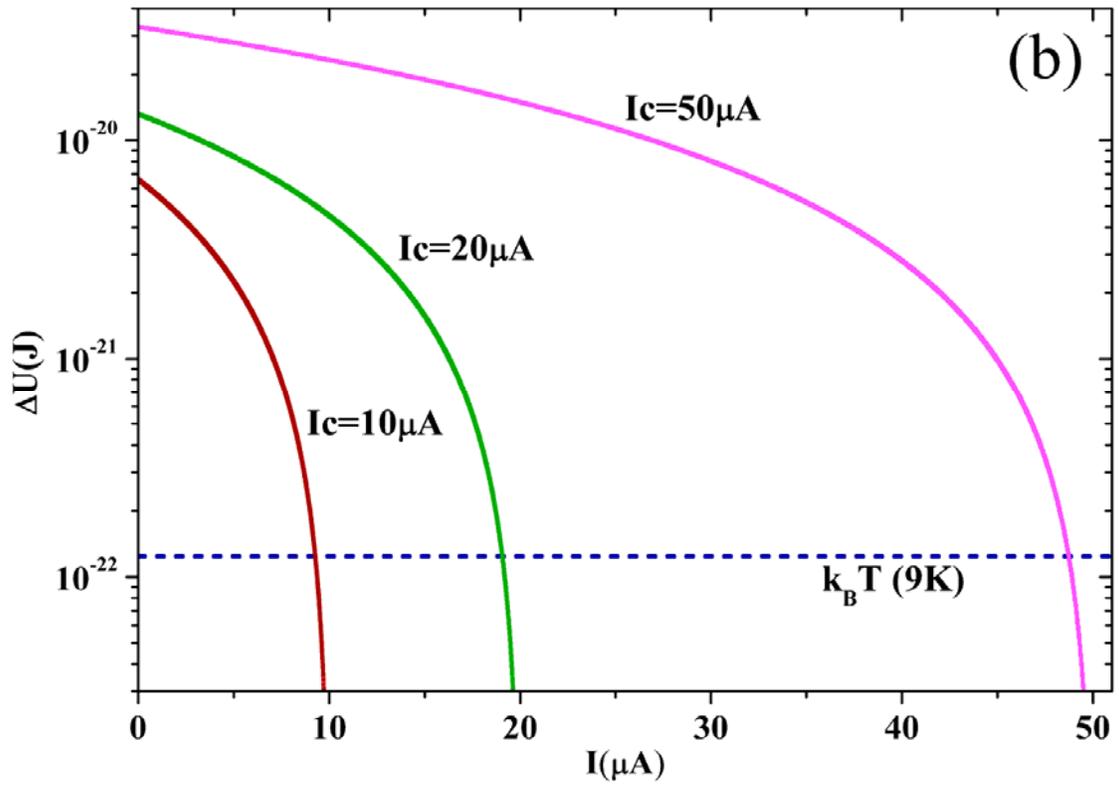

**I. Ottaviani et al. Figure 1**

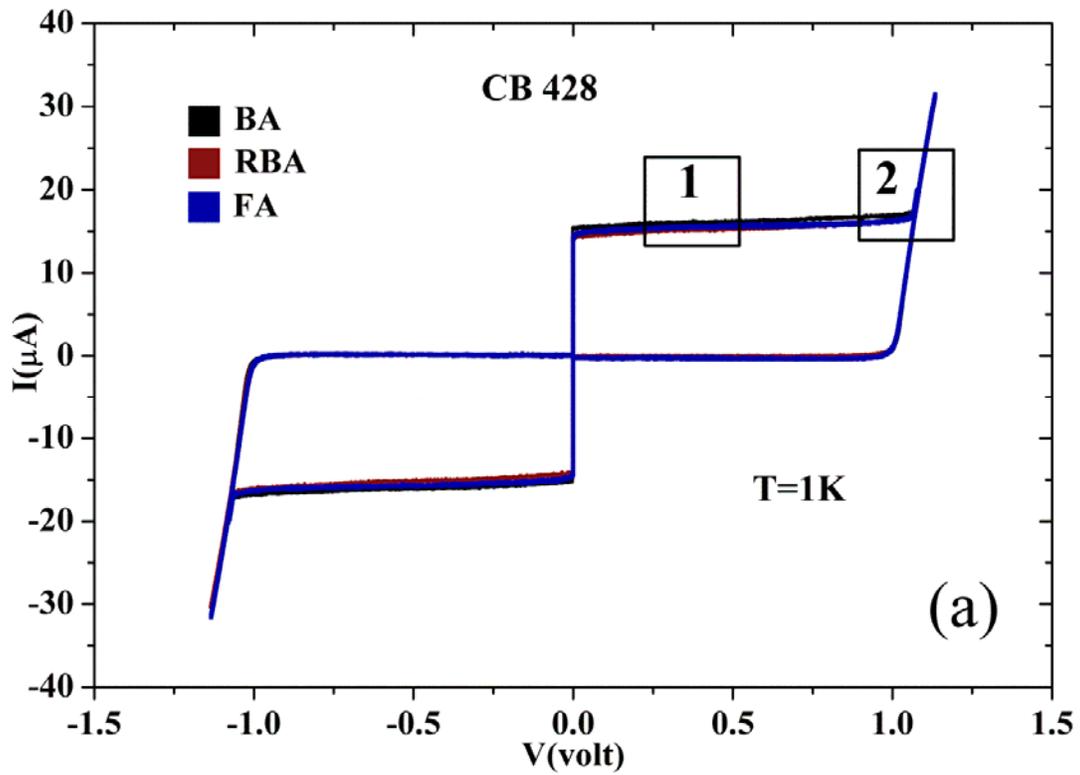

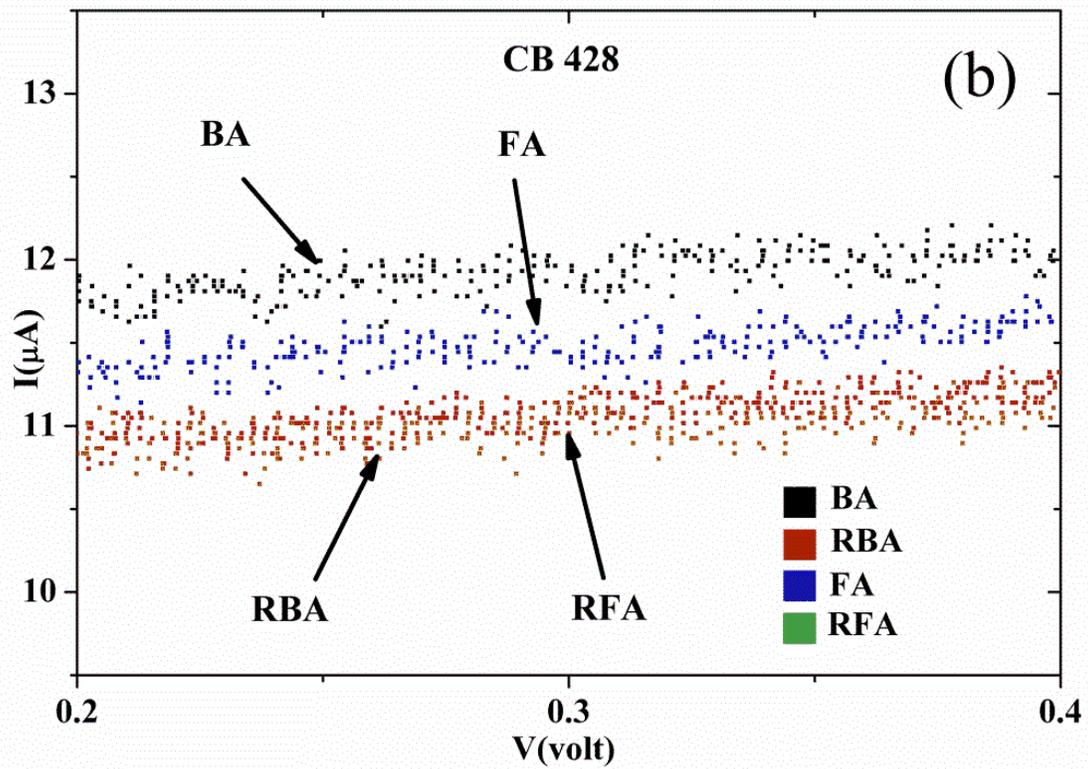

**I. Ottaviani et al. Figure 2 a,b**

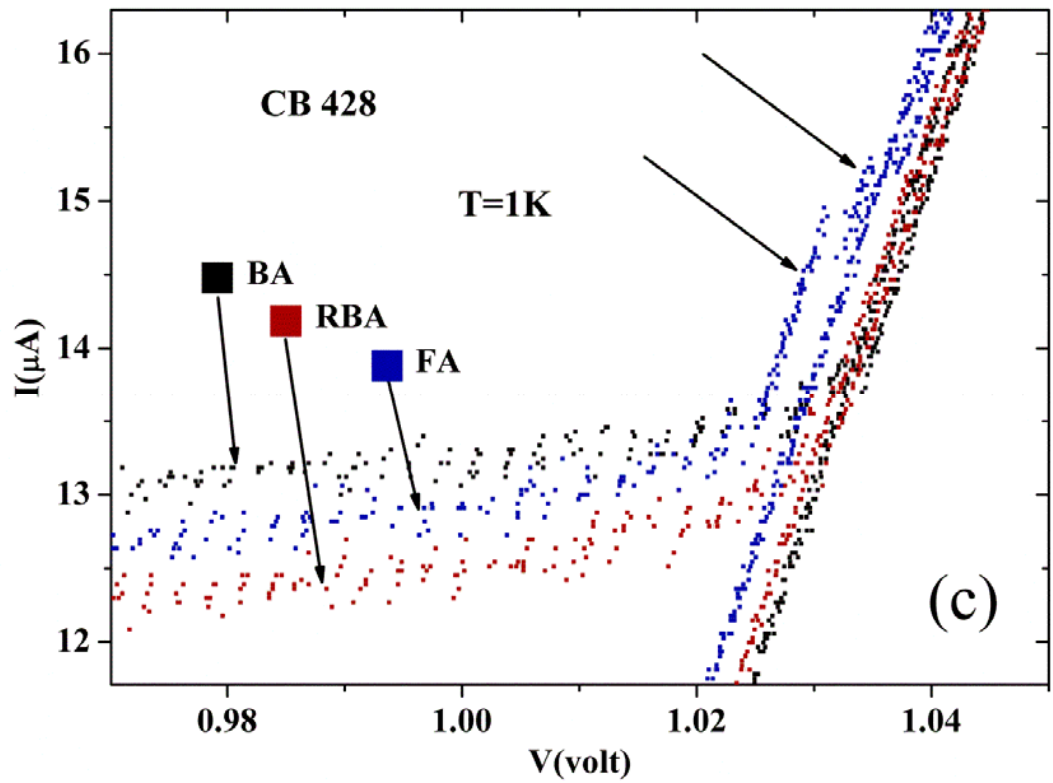
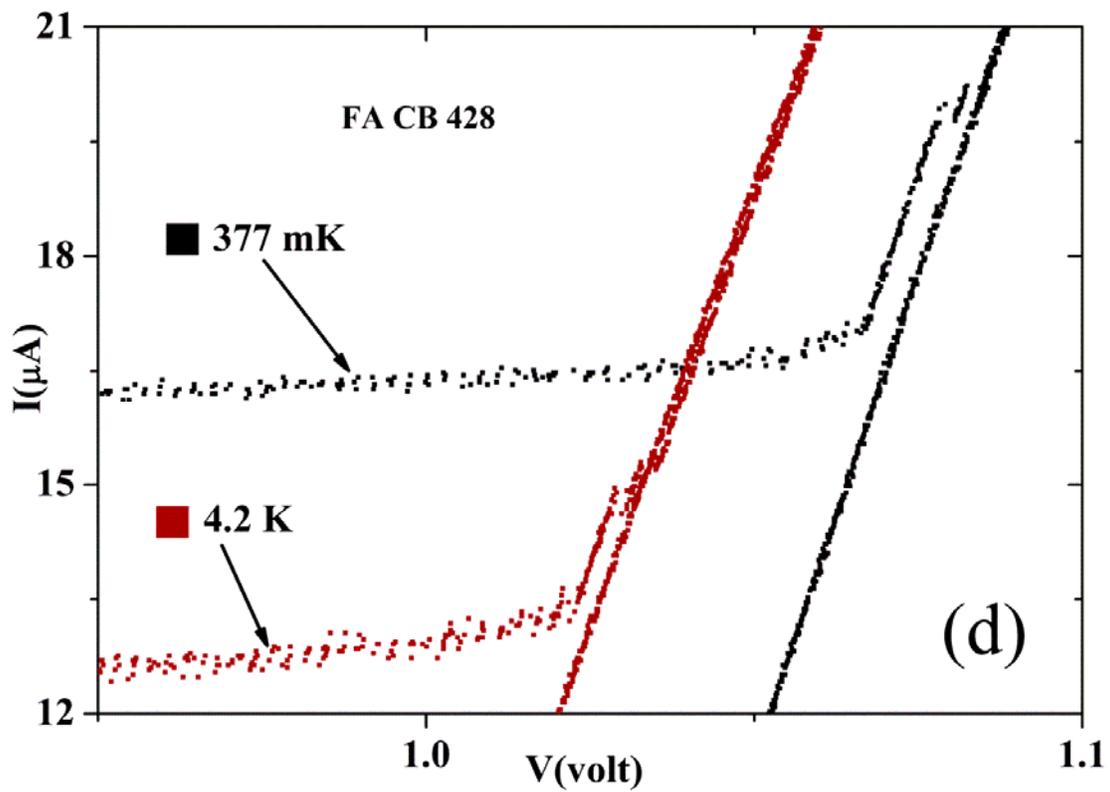

I. Ottaviani et al. Figure 2 c,d

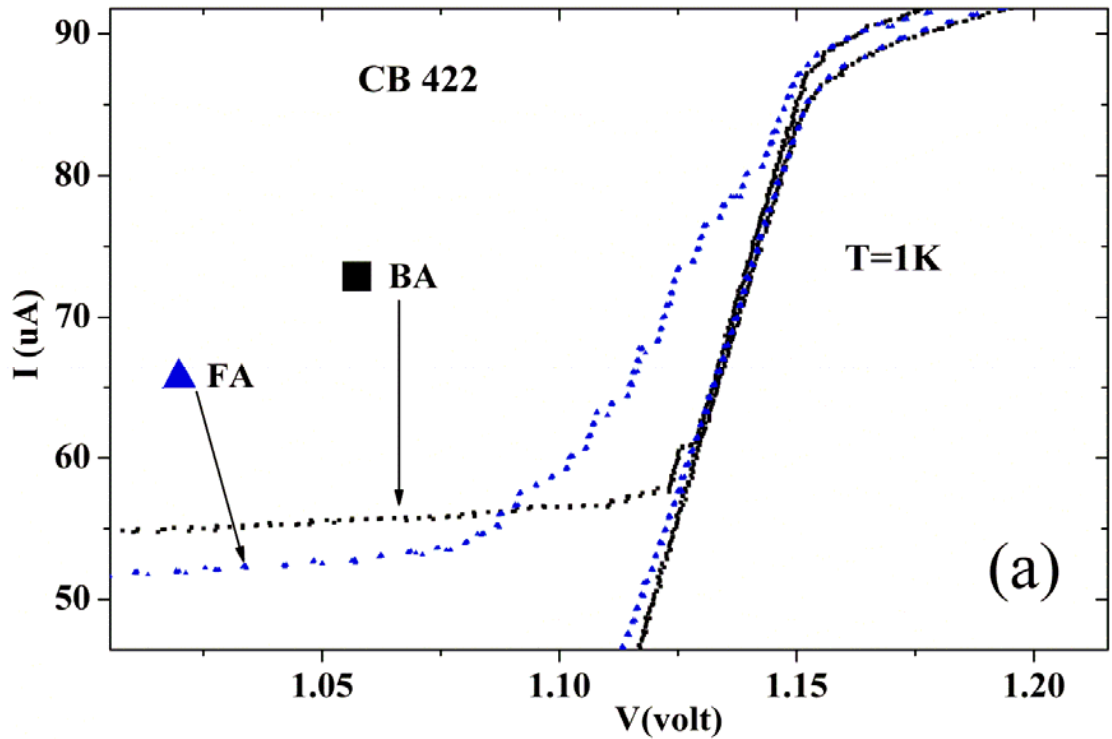

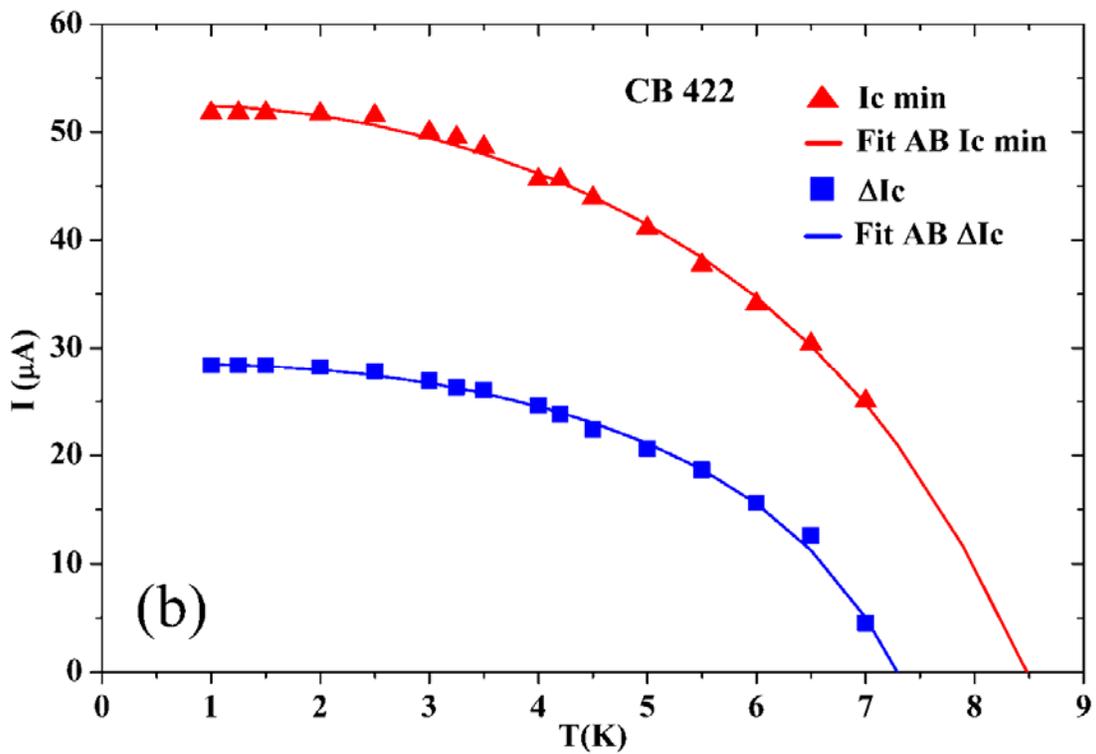

**I. Ottaviani et al. Figure 3**

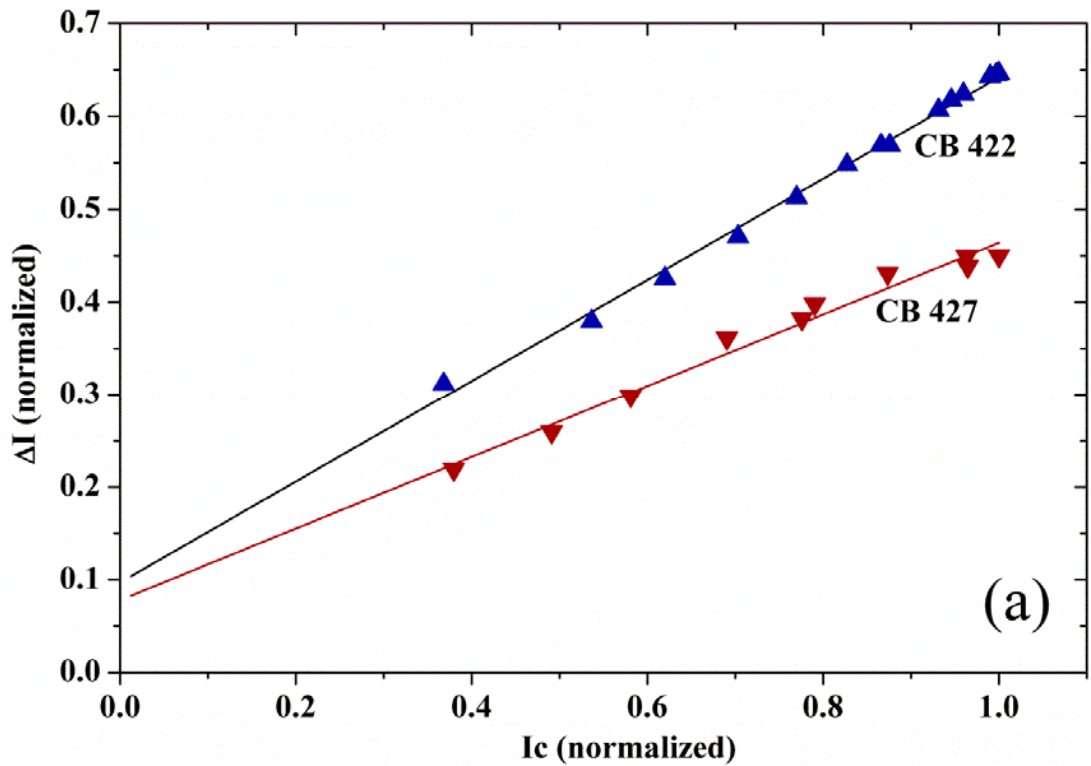
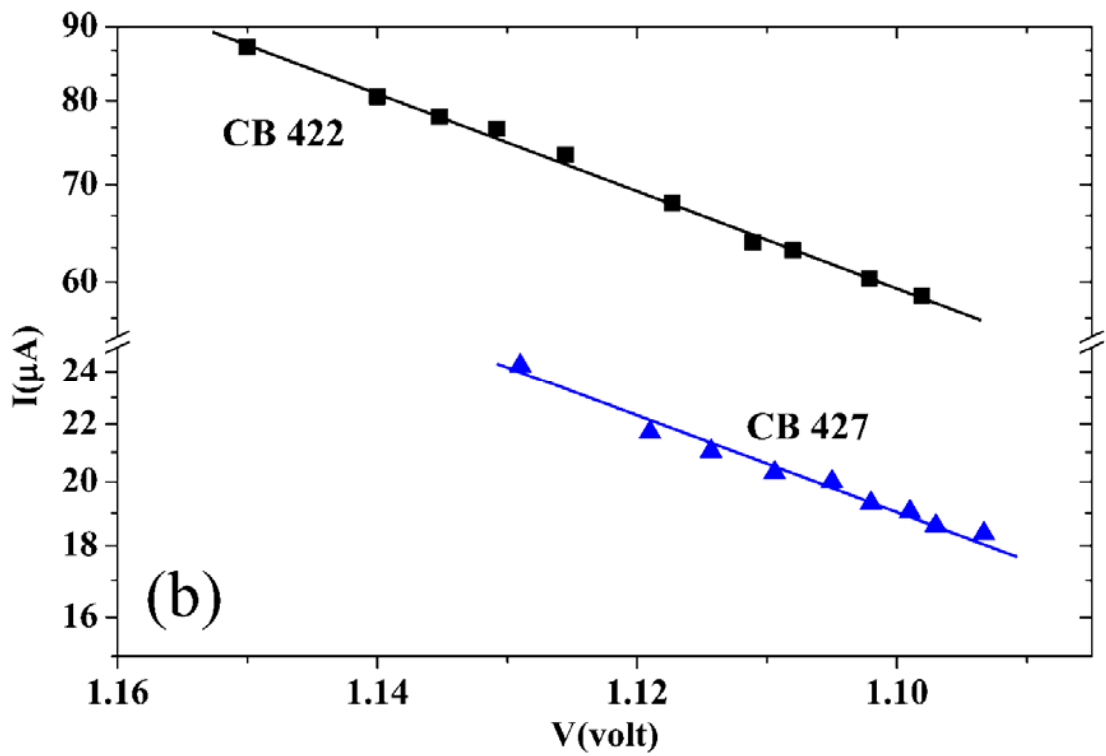

**I. Ottaviani et al. Figure 4**

15